\documentstyle[12pt]{article}
\title{\bf The Bose-Einstein distribution functions and the multiparticle
production at high energies}
\author{G.A. Kozlov\\
\em Bogoliubov Laboratory of Theoretical Physics,\\
\em Joint Institute for Nuclear Research,\\
\em 141980 Dubna, Moscow Region, Russia\\
\em e-mail: kozlov@thsun1.jinr.ru}

\begin{document}
\date{}
\maketitle
\begin{abstract}

{\small The evolution properties of propagating particles
produced at high energies in a randomly distributed environment
are studied. The finite size of the phase space of the multiparticle
production region as well as the chaoticity can be derived.\\

PACS number(s): 25.75.Gz, 12.40.Ee }
\end{abstract}

\section{Introduction}
\setcounter{equation}{0}

Interest in (charged) particles "moving" in an environment of
quantum fields taking into account the relations between quantum
fluctuations and chaoticity is still attracted by the
particle physics society. The events of high multiplicity at high
energies is an interesting subject because of
increasing energies of current and future colliders
 ($e^{+}e^{-}$, $\bar{p}p$, $pp$) of
particles. One of the most important tasks of superhigh energy
particle studies is to analyze fluctuations and correlations
such as the Bose-Einstein (BE) correlation [1] of produced
particles. This is a rather instructive tool to study high
multiplicity hadron processes in detail. We understand the
multiparticle (MP) production as the process of colliding particles
where the kinetic energy is dissipated into the mass of produced
particles [2]. We consider the incident energy
$\sqrt{s}\gg\Lambda $, where $\Lambda $ means the quantum
chromodynamics (QCD) scale. Phenomenological models [3,4]
describing the crucial properties of multiparticle correlations
are very useful for systematic investigations of the
properties caused by fluctuations and correlations. By
considering them, one can obtain the characteristic
properties of the internal structure of disordering of
produced particles in order to extract the information on the
space-time size of the multiparticle production region (MPPR), to
estimate the lifetime of the particle emitter, etc.
There were used the analysis of correlation functions (CF) and
distriburion functions (DF) in [5,6] to understand the possible
view on the quark-gluon plasma formation. In this paper, we
present the model to describe the very high
multiplicity effects at high energies. The most characteristic
point of our model is that the BE both DF's and CF's are taken into
account on the quantum level (the operators of production and
annihilations are used) with the random source contributions
coming from the environment. We define the average mean multiplicity
$\langle \bar{N}\rangle$ via MP CF $w(\vec{k})$ as $\langle \bar{N}\rangle
=\int d_{3}\,\vec{k}\,w(\vec{k})$, where $\vec{k}$ is the
spatial momentum of a particle. Following a natural way we
suppose $\langle \bar{N}\rangle\ll N$, while $N\ll N_{0}=\sqrt{s}/m$,
where $m\sim O(0.1~GeV)$. The main object in this investigation
is the MP thermal DF $\tilde{W}(k_{\mu},k^{\prime}_{\mu})$
related to $\langle N\rangle$ as
\begin{eqnarray}
\label{e1.1}
\tilde{W}(k_{\mu})= \langle N\rangle\,f(k_{\mu})
=\langle N\rangle {\langle b^{+}(k_{\mu})\,b(k_{\mu})\rangle
}_{\beta}\, ,
\end{eqnarray}
where $\langle N\rangle$ is defined as the scale of the
produced-particle multiplicity $N$ at four-momentum $k_{\mu}$
($\mu$ is the Lorenz index), the normalized DF $f(k_{\mu})$
is finite, i.e. $\int d_{4}k\,f(k_{\mu})<\infty$ and the label
$\beta$ in (\ref{e1.1})) means
the temperature T (of the phase space occupied by operators
$b^{+}(k_{\mu})$ and $b(k_{\mu}))$ inverse. The nature of operators
$b^{+}(k_{\mu})$ and $b(k_{\mu})$ will be clarified in Sec.3.

In this paper, we claim that the observation of the size effect
in a MP production is derived via MP CF's and DF's as well
as the so-called chaoticity which will be introduced in Sec.4.
The MP CF formalism concerns the statistical physics based on
the Langevin-type of equations. The Langevin equation to be
introduced in Sec.3 is considered as a basis for studying the
approach to equilibrium of the particle(s). It is assumed that
the heat bath being in essence infinite in size remains for all
times in equilibrium as well.
 We use the method applied to
the model where a relativistic particle moving in the Fock space
is described by the number representation underlying the second
quantization formulation of the canonical field theory. We deal
with the microscopic look at the problem with the elements of
quantum field theory (QFT) at the stochastic level with the
semiphenomenological noise embedded into the evolution
dissipative equation of motion.

\section{The model of dual representation within the dissipative dynamics}
\setcounter{equation}{0}

Let us suppose that the evolution of particles produced at high
energies is described by the solutions of the model Hamiltonian
where any physical system of particles is described by the
doublet of field operators. We introduce the dual states in the
eigenbasis $\{\vert k\rangle,\, \vert k(\epsilon)\rangle\}$ of
the Hamiltonian $H$, where $0\leq\epsilon <\infty$ is identified
as the frequency representing the energy of the object and $\vert
k\rangle$ is a discrete eigenstate. We consider the simple
dual model where H is given within the damped harmonic oscillator
\begin{eqnarray}
\label{e2.1}
H=H_{0}+H_{i}\ ,
\end{eqnarray}
where
\begin{eqnarray}
\label{e2.2}
H_{0}=\epsilon_{0}\vert k\rangle\langle
k\vert+\int^{\infty}_{0}d\epsilon\,\epsilon\,\vert k(\epsilon)\rangle
\langle k(\epsilon)\vert\ ,
\end{eqnarray}
\begin{eqnarray}
\label{e2.3}
H_{i}=\rho\int^{\infty}_{0}d\epsilon\,g(\epsilon)\, [\vert
k(\epsilon)\rangle\langle k\vert+\vert k\rangle\langle
k(\epsilon)\vert ]\ ,
\end{eqnarray}
$$ \langle k\vert k\rangle =1\, ,\,  \langle k\vert k(\epsilon)\rangle =
\langle k(\epsilon)\vert k\rangle =0\, ,\,\langle\epsilon\vert\epsilon^{\prime}
\rangle =\delta (\epsilon-\epsilon^{\prime})\, . $$
Here, we identify $\vert k\rangle$, $\vert k(\epsilon)\rangle$
and $\langle k\vert$, $\langle k(\epsilon)\vert$ with the
special mode operators of annihilation $a, b(\epsilon)$ and
creation $a^{+}, b^{+}(\epsilon)$, respectively, e.g., for
"quarks" and "gluons" or their combinations. In the interaction
Hamiltonian (\ref{e2.3}), $\rho$ is the coupling constant, while
$g(\epsilon)$ provides the transition between discrete and
continuous states.

The equations of motion obeying (\ref{e2.1}) with (\ref{e2.2}) and (\ref{e2.3})
are
$$i\,d_{t}a_{k}(t)=\epsilon_{0}\,a_{k}(t)+\rho\,\int^{\infty}_{0}d\epsilon\,
g(\epsilon)\,b_{k}(\epsilon,t)\, ,$$
$$i\,d_{t}b_{k}(\epsilon,t)=\epsilon\,b_{k}(\epsilon,t)+
\rho \,g(\epsilon)\,a_{k}(t)\, ,$$
where the label $k$ means $\vert\vec{k}\vert$ as the momentum. We
demand that $a_{k}(t)$ and $b_{k}(\epsilon,t)$ satisfy the
following natural conditions:
$$a_{k}(t)\,a^{+}_{k}(t)+\int^{\infty}_{0}d\epsilon\, b_{k}(\epsilon,t)\,
b^{+}_{k}(\epsilon,t) <\infty\, ,$$
$$ [a^{+}_{k}(t)]^{+}=a_{k}(t)\,\,\,\,
,[b^{+}_{k}(\epsilon,t)]^{+}=b_{k}(\epsilon,t)\, .$$
The next step is to give the relations between the variables and
couplings of the model (\ref{e2.1}) and the phenomenological
constants, e.g., the frequency $E$ and the damping constant
$\kappa$ involved into the dissipative dynamics given by the
Boltzmann-type equations in the relaxation time approximation:
\begin{eqnarray}
\label{e2.4}
\frac{d_{t}\langle a_{k}(t)\rangle }{\langle
a_{k}(t)\rangle}=-(i\,E+\kappa)\, ,
\end{eqnarray}
\begin{eqnarray}
\label{e2.5}
d_{t}w_{k}(t)=d_{t}\langle
a^{+}_{k}(t)a_{k}(t)\rangle=-2\,\kappa\,[w_{k}(t)-n(\beta)]\, .
\end{eqnarray}
Here,
$$d_{t}\langle a_{k}(t)\rangle =-i\,\epsilon_{0}\,\langle a_{k}(t)\rangle
-\rho^2\,\int^{\infty}_{0}\,d\epsilon\,g^2(\epsilon)\,\int^{t}_{\tau}\,ds\,
\langle a_{k}(s)\rangle\,\exp [-i\epsilon(t-s)]\, ,$$
$$\langle a_{k}(z)\rangle
=-i\,\rho\int^{z}_{\tau}\,dx\,\langle\sigma_{k}(x)\rangle\,\exp[-i\epsilon_{0}(z-x)]\,
,$$
$$\sigma_{k}(t)=\int^{\infty}_{0}\,d\epsilon\,g(\epsilon)\,b_{k}(\epsilon,t)\,
,$$
$$ n(\beta)=[\exp({\epsilon_{0}}\,\beta)\pm 1]^{-1}$$
as $\tau\rightarrow -\infty$. One can conclude that the
phenomenological constants $E$ and $\kappa$ in (\ref{e2.4}) (\ref{e2.5})
are nothing else but
$\epsilon_{0}$ and $2\,\pi\,\rho^2\,g^2(\epsilon_{0})$,
respectively, i.e. the microscopic parameters coming from the model
Hamiltonian (\ref{e2.1}).

\section{Stochastic model. Langevin equation}
\setcounter{equation}{0}
As it was pointed out in the Introduction, to derive the
characteristic features of the MP production physics at high
energies, one should specify the model on the quantal level.
Let us assume that only the particles $p_{i}$ of the same
kind of statistics labeled by index $i$ are produced just after
the high energy collision process occured, e.g., $pp,
\bar{p}p\rightarrow p_{i}$. In order to extend the method of
stochastically distributed particles in the environment, we
propose that rather complicated real physical processes happened in
the MP formation region should be replaced by a
single-constituent propagation of particles provided by a
special kernel operator (in the stochastically evolution
equation) considered as an input of the model and
disturbed by the random force $F$ [5,6]. We assume that $F$ can be the
external source being both a c-number function and an
operator. In such a hypothetical system of excited (thermal) local
phase we deal with the canonical operator $a(\vec{k}, t)$ and
its Hermitian conjugate $a^{+}(\vec{k}, t)$.
We formulate DF of produced particles in terms of a
point-to-point equal-time temperature-dependent thermal CF of two operators
\begin{eqnarray}
\label{e3.1}
w(\vec{k},\vec{k}^{\prime},t;T)=\langle a^{+}(\vec{k},t
)\,a(\vec{k}^{\prime},t)\rangle_{\beta} =
Tr [a^{+}(\vec{k},t)\,a(\vec{k}^{\prime},t)e^{-H\beta}]/Tr
(e^{-H\beta}).
\end{eqnarray}
 Here, $\langle ...\rangle_{\beta}$ means the procedure of thermal statistical
averaging; $\vec{k}$ and $t$ are, respectively, momentum and time variables,
$e^{-H\beta}/Tr(e^{-H\beta})$ stands for the standard density operator in
equilibrium, and the Hamiltonian $H$ is given by the squared form of the
annihilation $a_{p}$ and creation $a_{p}^{+}$ operators for Bose- and Fermi-
particles,
$H=\sum_{p}\epsilon_{p}a^{+}_{p}a_{p}$ (the energy $\epsilon_{p}$ and
operators $a_{p}$, $a_{p}^{+}$ carry some index $p$, where
$p_{\alpha}=2\pi\ n_{\alpha}/L, n_{\alpha}=0,\pm 1,\pm 2, ...; V=L^{3}$ is the
volume of the system considered).
Here, we use the canonical formalism in a stationary state in
the thermal equilibrium (SSTE), and a closed structural
resemblance between the SSTE and the standard QFT is revealed.
We define the thermal boson field as
\begin{eqnarray}
\label{e3.2}
\Phi_{B}(x_{\mu})=\frac{1}{\sqrt{2}}\left
[\phi(x_{\mu})+\phi^{+}(x_{\mu})\right ]\, ,
\end{eqnarray}
where
$$ \phi(x_{\mu})=\int\,d^{3}\vec{k}\,v_{k}\,a(\vec{k},t)\,\,  ,
v_{k}=\frac{e^{i\,\vec{k}\,\vec{x}}}{[(2\,\pi)^{3}\,\Delta(\vec{k})]^{1/2}}\, ,
$$
$$
\phi^{+}(x_{\mu})=\int\,d^{3}\vec{k}\,v^{+}_{k}\,a^{+}(\vec{k},t)\,\,
,
v^{+}_{k}=\frac{e^{-i\,\vec{k}\,\vec{x}}}{[(2\,\pi)^{3}\,\Delta(\vec{k})]^{1/2}}\,
$$
and $\Delta(\vec{k})$ is an element of the invariant phase volume.

     The standard canonical commutation relation (CCR)
\begin{eqnarray}
\label{e3.3}
{\left\lbrack a(\vec{k},t),a^{+}(\vec{k}^{\prime},t)
\right\rbrack}_{\pm}=\delta^{3}(\vec{k}-\vec{k}^{\prime})
\end{eqnarray}
at every time t is used as usual for Bose (-) and Fermi (+)-operators.

 The probability to find the particles in MPPR with momenta $\vec{k}$ and $\vec{k}^
{\prime}$ in the same event at the time $t$ is:
\begin{eqnarray}
\label{e3.4}
R(\vec{k},\vec{k}^{\prime},t)=W(\vec{k},\vec{k}^{\prime},t)/[W(\vec{k},t)
\cdot W(\vec{k}^{\prime},t)]\, ,
\end{eqnarray}
where the MP DF $W(\vec{k},t)$ in the simple version fluctuates
only its normalization, e.g., the mean multiplicity $\langle
N\rangle$ :

 Here, the one-particle thermal DF is defined as
$$W(\vec{k},t)=\langle N \rangle\cdot f(\vec{k},t)\ ,
$$
defining the single spectrum, while
$$W(\vec{k},\vec{k}^{\prime},t)=\langle N(N^{\prime}-\delta_{ij})
 \rangle\cdot f(\vec{k},\vec{k}^{\prime},t)\ $$
for $i$-and $j$-types of particles. Here,
$\delta_{ij}=1$ if $i=j$ and 0 otherwise. DF's $f(\vec{k},t)$ and
$f(\vec{k},\vec{k}^{\prime},t)$ look like (hereafter the label
$\beta$ will be omitted in the sense of (\ref{e1.1}) and
(\ref{e3.1}))
$$f(\vec{k},t)=\langle b^{+}(\vec{k},t)\, b(\vec{k},t)\rangle\, , $$
$$f(\vec{k},\vec{k}^{\prime},t)=\langle b^{+}(\vec{k},t)\ b^{+}(\vec{k}
^\prime,t)\ b(\vec{k},t)\ b(\vec{k}^{\prime},t)\rangle\ , $$
where
$$b(\vec{k},t)=a(\vec{k},t)+R(\vec{k},t) $$
 under the assumption of the random source-function $R(\vec{k},t)$
 being an operator, in general.
One can rewrite (\ref{e3.4}) in the following form
$$R(\vec{k},\vec{k}^{\prime},t)=\xi(N)\,\frac{f(\vec{k},\vec{k}^{\prime},t)}{
f(\vec{k},t)\,f(\vec{k}^{\prime},t)}\, ,$$
where
$$\xi(N)=\frac{\langle
N\,(N^{\prime}-\delta_{ij})\rangle}{\langle N\rangle\,\langle
N^{\prime}\rangle} \, .$$
For simplicity, we deal with operators $a$ and $ b$ as if they
are the single boson or fermion operators. Considering the
"propagation" of a particle with the momentum $\vec{k}$ in the
quantum equilibrium phase space under the infuence of a random
force coming from surrounding particles, the dissipative dynamics
of the relevant system is described by the equation containing
only the first order time derivatives of the dynamic
degrees of freedom, the operators $b(\vec{k},t)$ and $b^{+}(\vec{k},t)$
[5]:
\begin{eqnarray}
\label{e3.5}
i\ \partial_{t}b(\vec{k},t)=F(\vec{k},t)-A(\vec{k},t)+P\ ,
\end{eqnarray}
\begin{eqnarray}
\label{e3.6}
i\ \partial_{t}b^{+}(\vec{k},t)=A^{*}(\vec{k},t)-F^{+}(\vec{k},t)-P\
.
\end{eqnarray}
Here, the interaction of particles under consideration with
surroundings as well as providing the propagation is given by
the operator $A(\vec{k},t)$ defined as the one closely related
to the dissipation force:
\begin{eqnarray}
\label{e3.7}
A(\vec{k},t)=\int_{-\infty}^{+\infty}K(\vec{k},t-\tau)\ b(\vec{k},\tau)\ d\tau\ .
\end{eqnarray}
 The particle transitions are provided
by the random source operator $F(\vec{k},t)$ while $P$ stands
for a stationary external force.
 An interplay of particles with
surroundings is embedded into the interaction complex kernel
$K(\vec{k},t)$, while the real physical transitions are provided by the
random source operator $F(\vec{k},t)$ with the zeroth value of the statistical average
, $\langle F\rangle =0$.
The random evolution field operator $K(\vec{k},t)$ in (\ref{e3.7}) stands for
the random noise and it is assumed to vary stochastically with a $\delta$-
like equal time correlation function
$$\langle K^{+}(\vec{k},\tau)\ K(\vec{k}^\prime,\tau)\rangle =2\
{(\pi\rho)}^{1/2}\ \kappa\ \delta (\vec{k}-\vec{k}^\prime)\ ,
$$
where both the strength of the noise $\kappa$ and the positive
constant $\rho\rightarrow\infty$ define the effect of the
Gaussian noise on the evolution of particles in the
thermalized environment.

  The formal solutions of (\ref{e3.5}) and  (\ref{e3.6}) in the operator form
in $S(\Re_{4})$ ($k^{\mu}=(\omega=k^0,k_{j}))$ are
$$\tilde b(k_{\mu})=\tilde a(k_{\mu})+\tilde R(k_{\mu})\ , $$
$$\tilde b^{+}(k_{\mu})=\tilde a^{+}(k_{\mu})+\tilde R^{*}(k_{\mu})\ , $$
respectively, where the operator $\tilde a(k_{\mu})$ is expressed via the Fourier
transformed operator $\tilde F(k_{\mu})$ and the Fourier transformed kernel
function $\tilde K(k_{\mu})$ (coming from (\ref{e3.7})) as
$$\tilde a(k_{\mu})=\tilde F(k_{\mu})\cdot [\tilde K(k_{\mu})-\omega]^{-1}\
,$$
while the function $\tilde R(k_{\mu})$ $\sim P\cdot [\tilde K(k_{\mu})-\omega]^{-1}$.
In our model, we suppose that a heat bath (an environment) is an
assembly of damped oscillators coupled to the produced particles
which in turn are distributed by the random force $\tilde
F(k_{\mu})$. In addition, there is the assumption that the heat
bath is statistically distributed.
 The random
force operator $F(\vec{k},t)$ can be expanded by using the Fourier integral
\begin{eqnarray}
\label{e3.8}
F(\vec{k},t)=\int_{-\infty}^{+\infty}\frac{d\omega}{2\pi}\ \psi(k_{\mu})\
\hat c(k_{\mu})\ e^{-i\omega t}\ ,
\end{eqnarray}
where the form $\psi(k_{\mu})\cdot\hat c(k_{\mu})$ is just the Fourier
operator $\tilde F(k_{\mu})=\psi(k_{\mu})\cdot\hat c(k_{\mu})$, and the
canonical operator $\hat c(k_{\mu})$ obeys the commutation relation
$${\left\lbrack\hat c(k_{\mu}), \hat c^{+}(k_{\mu}^{\prime})\right\rbrack}_{\pm}=
\delta^{4}(k_{\mu}-k^{\prime}_{\mu})\ .$$
The function $\psi(k_{\mu})$ in (\ref{e3.8}) is determined by the
condition (the CCR (\ref{e3.3}) is taken into account)
$$\int_{-\infty}^{+\infty}\frac{d\omega}{2\pi}
{\left [\frac{\psi(k_{\mu})}{\tilde K(k_{\mu})-\omega}\right ]}^{2}=1\ . $$

\section{DF ratio enhancement.}
\setcounter{equation}{0}
  The enhanced probability for emission of identical
particles is given by the ratio $R$ of DF's in $S(\Re_{4})$ as follows:
\begin{eqnarray}
\label{e4.1}
R(k_{\mu},k_{\mu}^\prime;T)=\xi(N)\,\frac{\tilde f(k_{\mu},k_{\mu}^\prime;T)}
{\tilde f(k_{\mu})\cdot\tilde f(k_{\mu}^\prime)}\ ,
\end{eqnarray}
where $\tilde f(k_{\mu},k_{\mu}^\prime;T)=\langle\tilde b^{+}(k_{\mu})\ \tilde b^
{+}(k_{\mu}^\prime)\ \tilde b(k_{\mu})\ \tilde b(k_{\mu}^\prime)\rangle
$ and $\tilde f(k_{\mu})=\langle\tilde b^{+}(k_{\mu})\ \tilde b(k_{\mu})\rangle
$.
 Using Fourier solutions of equations (\ref{e3.5}) and (\ref{e3.6}) in $S(\Re_{
4})$, one can get the R-ratio for DF
\begin{eqnarray}
\label{e4.2}
R(k_{\mu},k_{\mu}^\prime;T)=\xi(N)\,[1+D(k_{\mu},k_{\mu}^\prime;T)]\,
,
\end{eqnarray}
where
\begin{eqnarray}
\label{e4.3}
D(k_{\mu},k_{\mu}^\prime;T)=\frac{\Xi(k_{\mu},k_{\mu}^\prime)[\Xi(k_{\mu}^
\prime,k_{\mu})+\tilde R^{+}(k_{\mu}^\prime)\tilde R(k_{\mu})]+
\Xi(k_{\mu}
^\prime,k_{\mu})\tilde R^{+}(k_{\mu})\tilde R(k_{\mu}^\prime)}{
\tilde f(k_{\mu})\cdot\tilde f(k_{\mu}^\prime)}
\end{eqnarray}
and the Bose-Einstein CF $\Xi(k_{\mu},k_{\mu}^\prime)$ looks like
\begin{eqnarray}
\label{e4.4}
\Xi(k_{\mu},k_{\mu}^\prime)=\langle\tilde a^{+}(k_{\mu})\ \tilde a(k_{\mu}^
\prime)\rangle \cr
=\frac{\psi^{*}(k_{\mu})\cdot\psi(k_{\mu}^\prime)}{[\tilde K^{*}(k_{\mu})-
\omega]\cdot[\tilde K(k_{\mu}^\prime)-\omega^\prime]}\cdot\langle\hat c^{+}
(k_{\mu})\ \hat c(k_{\mu}^\prime)\rangle\ .
\end{eqnarray}
Inserting CF (\ref{e4.4}) into (\ref{e4.3}) and taking into
account the trick with $\delta^{4}(k_{\mu}-k_{\mu}^{\prime})$-function
to be replaced by the $\delta$-like consequence like $\Omega(r)\cdot\exp
[-(k-k^{\prime})^{2}r^{2}]$ [7], one can get the following
expression for D-function instead of (\ref{e4.3})

\begin{eqnarray}
\label{e4.5}
D(k_{\mu},k_{\mu}^\prime;T)= \lambda(k_
{\mu},k_{\mu}^\prime;T)\cdot\exp(-q^2/2) \cr
\times [n(\bar{\omega},T)\Omega(r) \exp(
-q^2/2)+\tilde R^{*}(k_{\mu}^\prime)\tilde R(k_{\mu})+\tilde R^{*}(k_{
\mu})\tilde R(k_{\mu}^\prime)]\ ,
\end{eqnarray}
where
$$\lambda(k_{\mu},k_{\mu}^\prime;T)=\frac{\Omega(r)}{\tilde f(k_{\mu})\cdot
\tilde f(k_{\mu}^\prime)}\cdot n(\bar{\omega},T)\ ,\ \bar{\omega}=\frac{1}{2}
(\omega+\omega^\prime)\ ,$$
while $ q^{2}\equiv Q^{2}\,r^{2}$ and
the function $\Omega(r)\cdot n(\omega;T)\cdot \exp(-q^2/2)$ in (\ref{e4.5})
describes the
space-time size of the multiparticle production region.
 Choosing the z-axis along the $pp$ or $\bar{p}p$ collision axis one can put
$$Q_{\mu}=(k-k^\prime)_{\mu}, Q_{0}=\epsilon_
{\vec{k}}-\epsilon_{\vec{k}^\prime}, Q_{z}=k_{z}-k_{z}^\prime, Q_{t}=
{{[(k_{x}-k_{x}^\prime)}^2+{(k_{y}-k_{y}^\prime)}^2]}^{1/2}\ , $$
$$\Omega(r)=\frac{1}{\pi^2}\, r_{0}\cdot r_{z}\cdot r_{t}^2\ ,$$
where $r_{0}$, $r_{z}$ and $r_{t}$ are time-like, longitudinal, and transverse
"size" components of MPPR.
To derive (\ref{e4.5}), the  Kubo-Martin-Schwinger condition
 ( $\mu$ is the chemical potential)
$$\langle a(\vec{k}^\prime,t^\prime)\ a^{+}(\vec{k},t)\rangle = \langle a^{+}
(\vec{k},t)\ a(\vec{k}^\prime,t-i\beta)\rangle\cdot \exp(-\beta\ \mu)\
$$
has been used, and
the thermal statistical averages for the $\hat c(k_{\mu})$-operator
should be represented in the following form:
\begin{eqnarray}
\label{e4.6}
\langle\hat c^{+}(k_{\mu})\ \hat c(k_{\mu}^\prime)\rangle =\delta^{4}(k_{\mu}-
k_{\mu}^\prime)\cdot n(\omega,T)\ ,
\end{eqnarray}
\begin{eqnarray}
\label{e4.7}
\langle\hat c(k_{\mu})\ \hat c^{+}(k_{\mu}^\prime)\rangle =\delta^{4}(k_{\mu}-
k_{\mu}^\prime)\cdot [1\pm n(\omega,T)]
\end{eqnarray}
for Bose (+)- and Fermi (-)-statistics, where $n(\omega,T)=\{\exp[(\omega-\mu)
\beta]\pm 1\}^{-1}$.
Formula (\ref{e4.5}) indicates that the chaotic multiparticle
source emanating from the thermalized MPPR exists.
Taking into account (\ref{e4.6}) and (\ref{e4.7}), it is easy
to see that the correlation functions containing the random
force functions $F(\vec{k},t)$ (\ref{e3.8}) carry the quantum
features in the termalized stationary equilibrium, namely
$$\langle
F(\vec{k},t)\,F^{+}(\vec{k^{\prime}},t^{\prime})\rangle=\delta^{3}(\vec{k}-\vec
{k}^{\prime}))\,\Gamma_{1} (\vec{k},-\Delta t)\, ,$$
$$\Gamma_{1} (\vec{k},-\Delta
t)=\int\frac{d\omega}{2\,\pi}\,\vert\psi(k_{\mu})\vert^2\,[1\pm
n(\omega, \beta)]\,\exp (-i\,\omega\,\Delta t)\, ,$$
$$\Delta t= t-t^{\prime}\, ;$$
$$\langle
F^{+}(\vec{k},t)\,F^{+}(\vec{k^{\prime}},t^{\prime})\rangle=\delta^{3}(\vec{k}-\vec
{k}^{\prime}))\,\Gamma_{2} (\vec{k},\Delta t)\, ,$$
$$\Gamma_{2} (\vec{k},\Delta
t)=\int\frac{d\omega}{2\,\pi}\,\vert\psi(k_{\mu})\vert^2\,
n(\omega, \beta)\,\exp (i\,\omega\,\Delta t)\, .$$
The quantitative information (longitudinal $r_{z}$ and transverse $r_{t}$
components of MPPR, the temperature T of the environment)
could be extracted by fitting the theoretical formula (\ref{e4.5}) to
the measured D-function and estimating the errors of the fit parameters.
 Hence, the measurement of the space-time
evolution of the multiparticle source would provide information of the
multiparticle process
 and the general reaction mechanism. The
temperature of the environment enters into formula (\ref{e4.5}) through the
two-particle CF $\Xi(k_{\mu},k_{\mu}^\prime;T)$. If T is unstable, the
$R$-functions (\ref{e4.1}) will change due to a change of
DF $\tilde f$ which, in fact, can be considered as an effective density
of the  multiparticle source.
   Formula (\ref{e4.1}) looks like the
fitting R-ratio using a source parametrization:
$$R_{F}(r)=const\,[1+\lambda_{F}(r)\cdot \exp(-{r_{t}^2\cdot Q_{t}^2}/2 -{r_{z}^2
\cdot Q_{z}^2}/2)]\ , $$
where $r_{t}(r_{z})$ is the transverse (longitudinal) radius parameter of the
source with respect to the beam axis, $\lambda_{F}$ stands for the effective
intercept parameter (chaoticity parameter) which has a general dependence of
the mean momentum of the observed particle pair. Here, the dependence on the
source lifetime is omitted.
 The chaocity parameter
$\lambda_{F}$ is the temperature-dependent and the positive one defined by
$$\lambda_{F}(r)=\frac{{\vert\Omega(r)\
n(\bar{\omega};T)\vert }^2}{\tilde f(k_{\mu})\cdot\tilde f(k_{\mu}^\prime)
}\ .$$

   Comparing (\ref{e4.3}) and (\ref{e4.4}) one can identify
$$\Xi(k_{\mu},k_{\mu}^\prime)=\Omega(r)\cdot n(\bar{\omega};T)
\cdot \exp(-q^2/2)\ .$$
Hence, CF $\Xi(k_{\mu},k_{\mu}^\prime)$ defines uniquely the
size $r$ of MPPR.
There is no  satisfactory tool to derive the precise analytic
form of the random source function $\tilde R(k_{\mu})$ in
(\ref{e4.3}), but one can put (see (\ref{e4.4}) and taking into
account $\tilde R(k_{\mu})$ $\sim P\cdot [\tilde
K(k_{\mu})-\omega]^{-1}$) [8,5,6] that
$$\tilde R(k_{\mu})={[\alpha\cdot \Xi(k_{\mu})]}^{1/2}\ , $$
where $\alpha$ is of the order $O\left
(P^2/n(\omega,T)\cdot{\vert\psi(k_{\mu})\vert}^2\right )$.
 Thus,
\begin{eqnarray}
\label{e4.8}
D(q^2;T)=\frac{\tilde\lambda^{1/2}(\bar{\omega};T)}{(1+\alpha)(1+
\alpha^\prime)}e^{-q^{2}/2}\left [\tilde\lambda^{1/2}(\bar{\omega};T)
e^{-q^{2}/2}+2{(\alpha\alpha^\prime)}^{1/2}\right]\ ,
\end{eqnarray}
where
$$\tilde\lambda(\bar{\omega};T)=\frac{n^{2}(\bar\omega;T)}
{n(\omega;T)\cdot n(\omega^{\prime};T)}\ .$$
It is easy to see that, in the vicinity of $q^{2}\approx 0$, one can get
the full correlation if $\alpha =\alpha^\prime =0$ and $\tilde\lambda (\bar
{\omega};T)$=1. Putting $\alpha =\alpha^\prime$ in (\ref{e4.8}), we find
the formal lower
bound on the space-time dimensionless size of MPPR of the bosons
$$q^2\geq\ln\frac{\tilde\lambda (\bar{\omega};T)}{{[\sqrt{(\alpha+1)^{2}+
\alpha^2}-\alpha]}^2}\ .$$
 In fact, the function $D(k_{\mu},k_{\mu}^\prime;T)$ in (\ref{e4.5})
could not be observed because of some
model uncertainties. In the real world, D-function has to contain
background contributions
which have not been included in the calculation.
To be close to the experimental data, one has to expand the $D$-
function as projected on some well-defined function (in $S(\Re_{4})$)
of the relative momentum of bosons produced
$D(k_{\mu},k_{\mu}^\prime;T)\rightarrow D(Q_{\mu}^{2};T).$
Thus, it will be very instructive to use the polynomial expansion which is
suitable to avoid any uncertainties as well as to characterize the degree of
deviation from the Gaussian distribution, for example. In $(-\infty,+\infty)$,
a complete orthogonal set of functions can be obtained with the help of the
Hermite polynomials in the Hilbert space of the square integrable functions
with the measure $d\mu(z)=\exp(-z^2/2)dz$. The function $D$ corresponds to
this class if
$$\int_{-\infty}^{+\infty}dq \exp(-q^2/2)\ \vert D(q)\vert^n\ <\infty\ ,
 n=0,1,2,...\ .$$
The expansion in terms of the Hermite polynomials $H_{n}(q)$
\begin{eqnarray}
\label{e4.9}
D(q)=\lambda\sum_{n}\ c_{n}\cdot H_{n}(q)\cdot \exp(-q^2/2)\
\end{eqnarray}
is well suited for the study of possible deviations both from the experimental
shape and from the exact theoretical form of the function $D$
(\ref{e4.5}). The coefficients $c_{n}$ in (\ref{e4.9}) are defined via
the integrals over
the expanded functions $D$ because of the orthogonality condition
$$\int_{-\infty}^{+\infty} H_{n}(x)\ H_{m}(x)\ \exp(-x^2/2)\ dx=\delta_{n,m}
\ .$$
Thus, the observation of the multiparticle correlation enables one
 to extract the properties of the structure of
$q^2$, i.e. the space-time size of MMPR.
The other possibility is related with the replacement of
$R$- function (\ref{e4.2}) with respect to the
cylindrical symmetry angles $\theta$ and $\phi$ which are
non-observable ones at fixed $Q_{t}$:
$$
R(k_{\mu},k_{\mu}^{\prime};T)\rightarrow\bar{R}(Q_{t};T)=
C_{N}^{-1}\,\xi(N)\,\int dq_{t}\,dQ_{z}\,d\theta\, d\phi\,\tilde
{f}(k_{\mu},k_{\mu}^{\prime};T)\ ,$$
where
$$ C_{N}=\int dq_{t}\,dQ_{z}\,d\theta\, d\phi\,\tilde
{f}(k_{\mu})\,\tilde{f}(k_{\mu}^{\prime})\, ,$$
$$q_{t}=\frac{1}{\cos\theta+\sin\theta}
\,\left\{k_{x}+k_{y}\mp\frac{1}{2}Q_{t}\left [\cos (\theta+\phi)+\sin
(\theta+\phi)\right ]\right\}\ .$$
Then, $\bar{R}(Q_{t};T)=\xi(N)\,[1+\bar{D}(Q_{t};T)]$ with
$$\bar{D}(Q_{t};T)=\frac{\bar{C_{N}}^{-1}(T)}{(1+\alpha)(1+\alpha^{\prime})}\,
\exp [-(r^{2}_{t}\,Q^{2}_{t})]\,F(Q_{t};T)\, ,$$
$$F(Q_{t};T)=\int dq_{t}\,dQ_{z}\,d\theta\,
d\phi\,n^{2}(\bar{\omega};T)\,e^{-\beta_{0z}}[1+2\sqrt{\alpha\alpha^{\prime}
\tilde{\lambda}^{-1}(\bar{\omega};T)}\,e^{q^{2}/2}]\, ,$$
$$\beta_{0z}\equiv r^{2}_{0}\,Q^{2}_{0}+r^{2}_{z}\,Q^{2}_{z}\, ,$$
$$\bar{C_{N}}(T)=\int dq_{t}\,dQ_{z}\,d\theta\,
d\phi\,n(\omega;T)\,n(\omega^{\prime};T)\,.$$

It remains to sketch how one goes about calculating the
thermodynamical quantities in a local thermalized system of
produced particles. Taking into account the positive- and
negative-frequency parts of the boson field operator (\ref{e3.2}) to
be applied to the energy-momentum tensor
$T_{\mu\nu}(x)=:\Phi_{B}(x)\,\tilde{\vec{k_{\mu}}}\tilde{\vec{k_{\nu}}}\,\Phi_{B}(x):$
and the particle flow operator
 $\Pi_{\mu}(x)=:\Phi_{B}(x)\,\tilde{\vec{k_{\mu}}}\,\Phi_{B}(x):$
we can calculate the energy density $E(\beta)$, the
pressure $V(\beta)$ and the entropy density $S(\beta)$ in the
local system of the volume $v$ for bosons in the equilibrium thermalized phase
space. The simple straightforward calculations give (see also [9])
$$E(\beta)=\frac{1}{(2\,\pi)^2\,v}\int\,d_{3}\vec{k}\,d\omega\,\omega^{2}\,
M(k_{\mu},\beta)\, ,$$
$$V(\beta)=\frac{1}{6\,\pi^2\,v}\int\,d_{3}\vec{k}\,\vec{k}^2\,d\omega\,
M(k_{\mu},\beta)\, ,$$
$$S(\beta)=\frac{1}{2\,\pi\,v}\int\,d_{3}\vec{k}\, \{
[1+\Pi_{0}(\vec{k},\beta)]\,\ln[1+\Pi_{0}(\vec{k},\beta)]-\Pi_{0}(\vec{k},\beta)\,
\ln \Pi_{0}(\vec{k},\beta)\}\, ,$$
where
$$\Pi_{0}(\vec{k},\beta)=\frac{1}{2}\,\int^{+\infty}_{-\infty}\,d^{2}\omega\,
M(k_{\mu},\beta)\, ,$$
$$M(k_{\mu},\beta)=\frac{\psi^{2}(k_{\mu})}{\vert\tilde{K}(k_{\mu})-\omega\vert^2}\cdot
n(\omega,\beta)\,  ,$$
$$d_{3}\vec{k}\equiv\frac{d^{3}\vec{k}}{\sqrt{(2\,\pi)^3\,\Delta(\vec{k})}}\,
$$
$$\Phi_{B}(x)\,\tilde{\vec{k_{\mu}}}\,\Phi_{B}(x)\equiv\frac{1}{2}
[\Phi_{B}(x)\,(k_{\mu}\Phi_{B}(x))
-(k_{\mu}\Phi_{B}(x))\,\Phi_{B}(x)]\, .$$
Here, we suppose that the thermalized MPPR is isotropic, and one can
use the space-averaged operators normalized to the volume $v$,
taking ensemble averages (\ref{e4.6}) and  (\ref{e4.7}). It
is easy to see that both $E(\beta)$ and  $V(\beta)$ tend to
their maximum values with rising $T$, while the entropy $S(\beta)$
changes not so much essentially even if $T\rightarrow\infty$.
\section{Statistical distributions}
From a widely accepted point of view, at high energies, there
are two channels, at least, for multiparticle production where
produced particles occupy the MMPR consisting of $i$ elementary
cells. These main channels are a) a direct channel assuming that
all particles $p_{j}$ are produced directly within the quark
($q$)-antiquark ($\bar{q}$) annihilation or the gauge-boson
fusion, e.g., $q\bar{q}\rightarrow p_{j}p_{j}...$; b) an indirect
channel  which means that the particles are produced via the
decays of intermediate vector bosons $\chi^{*}$ in both heavy
and light sectors in the kinematically allowed region, e.g.,
$q\bar{q}\rightarrow\chi^{*} \chi^{*}...\rightarrow
p_{j}p_{j}...$ All the particles produced are classified by the
like-sign constituents that are labeled as $p^{+}\, ,p^{-}\,
,p^{0}$-subsystems, where $p :\mu\, ,\pi \, ,K...$
The mean multiplicity $\langle N\rangle$ and the mean energy $\langle E\rangle$
of the $p_{j}$ subsystem are defined as [3]
$$\langle
N\rangle=\sum_{j}\,\sum_{m_{j}}\,m_{j}\,\zeta^{(m_{j})}_{j}\,
,$$
$$\langle
E\rangle=\sum_{j}\,\sum_{m_{j}}\,m_{j}\,\epsilon_{j}\,\zeta^{(m_{j})}_{j}\,
,$$
where $\epsilon_{j}$ is the energy of a $p$-particle in the $j$th
elementary cell and $\zeta^{(m_{j})}_{j}$ stands for the probability of
finding $m_{j}$ $p$ particles in the $j$th cell and is
normalized as
$$\sum_{m_{j}=0}^{\infty}\,\zeta^{(m_{j})}_{j}=1\, .$$
In the direct channel, for charged produced mesons
$\langle N\rangle$ is defined uniquely for a given $\beta$ as
$$\langle N\rangle=2\,\sum_{j}\,[\exp(\epsilon_{j}\,\beta)-1]^{-1}\,  ,$$
while $\langle E\rangle$ is
$$\langle
E\rangle=\frac{1}{3}\,\sqrt{s}=\sum_{j}\frac{\epsilon_{j}}
{\exp(\epsilon_{j}\,\beta)-1}\,  .$$
Going into $y$-rapidity space in the longitudinal phase space
with a lot of cells of equal size $\delta y$, the energy $\epsilon_{j}$
should be expressed in terms of the transverse mass
$m_{t}=\sqrt{{\langle{k_{t}}\rangle}^2+m^{2}_{p}}$
($\langle{k_{t}}\rangle$ and $m_{p}$ are the transverse average
momentum and the mass of a $p$-particle):
$$\epsilon_{j}(s)=\frac{m_{t}}{2}\,[g_{j}(\tilde{s})+g^{-1}_{j}(\tilde{s})]\,\,
   ,\tilde{s}=\frac{s}{4\,m^{2}_{t}}\, ,$$
$$g_{j}(\tilde{s})=(\sqrt{\tilde{s}}+\sqrt{\tilde{s}-1})\,\exp
[-(j-1/2)\,\delta y]\, .$$
Here, the four-momentum of $p$-particle is given as
$$k^{\mu}=(\sqrt{{\langle{k_{t}}\rangle}^2+m^{2}_{p}}\,
\cosh y, k_{t}\,\cos\varphi , k_{t}\,\sin\varphi ,
\sqrt{{\langle{k_{t}}\rangle}^2+m^{2}_{p}}\,\sinh y)\,  ,$$
where the azimuthal angle of $k_{t}$ is in the range
$0<\varphi <2\,\pi$.
Our model produces an enhancement of $R(Q,\beta)$ in the small
enough region of $Q$ where $R$ is defined only by the model
parameter $\alpha$ and the mean multiplicity $\langle
N(s)\rangle$ at fixed value of $\beta$, namely:
$$R(Q,\beta)\simeq\xi(\langle N\rangle)\left\{1+
\frac{\sqrt{\tilde\lambda(\bar{\omega},\beta)}}
{(1+\alpha)^2}
\left [\sqrt{\tilde\lambda(\bar{\omega},\beta)}+2\,\alpha -
(\sqrt{\tilde\lambda(\bar{\omega},\beta)} +\alpha)
Q^2\,r^2\right ]\right\}\, .$$
It is clear that $R(Q,\beta)$-function at $Q^2=0$
$$R(Q,\beta)\simeq\xi(\langle N(s)\rangle)\cdot\left [2-\left
(\frac{\alpha}{1+\alpha}\right )^2\right ] $$
cannot exceed 2 because of $\alpha\neq 0$ and $\xi(N(s))< 1$
even at large multiplicity. The Boltzmann behaviour should be
realized in the case when $\alpha\rightarrow\infty$, i.e. the
main contribution to the fluctuating behaviour of the $R(Q,\beta)$-function
should come from the random source contribution
(see (\ref{e4.5}) and (\ref{e4.8})). We found that the
enhancement of the $R(Q,\beta)$-function, mainly, the shape of
this function, strongly depends on the transverse size $r_{t}$
of the phase space and has a very weak dependence of the $\delta y$ size of
a separate elementary cell. The increasing of $r_{t}$ makes
that the shape of the $R(Q,\beta)$-function becomes more
crucial.

Obviously, $\xi(\langle N(s)\rangle)$ is the normalization
constant in (\ref{e4.2}), where
$\langle N(s)\rangle$ should be derived at the origion of $Q^2$
precisely from $R(Q=0,\beta)\equiv R_{0}(s)$ as
$$\langle N(s)\rangle\simeq\frac{1}{\varepsilon}\,  ,$$
where
$$\varepsilon=1-\frac{R_{0}(s)}{2-\left
(\frac{\alpha}{1+\alpha}\right )^2}\, $$
can be extracted from the experiment at some chosen value of
$\alpha$ ($\alpha=0$ should be taken into account as well).
On the other hand, the $R(Q,\beta)$-function allows one to
measure $\alpha=\alpha^{\prime}$ which parametrizes the random
source contribution as well as the splitting between $\alpha$
and $\alpha^{\prime}$. Neglecting the random source contribution
(i.e., putting $\alpha=\alpha^{\prime}=0$) we can estimate the
chaoticity $\tilde\lambda (\bar{\omega},\beta)$ by measuring
$R(Q,\beta)$ as $Q^2\rightarrow 0$.

In fact, the theoretical prediction that $D(Q,\beta)>1$
 means that in MPPR one should  select the single
boson "dressing" of some quantum numbers, and the particles
suited near it in the phase space are "dressed" with the same
set of quantum numbers. The amount of such neighbour particles
has to be as many as possible. This allows to form a cell in the
space-time occupied by equal-statistics particles only. Such
a procedure can be repeated while all the particles will occur
in the MPPR. This leads to the space-time BE distribution of
produced particles in the phase-space cells formed only for
bosons. In fact, there is no restriction of the number of bosons
occupying the chosen elementary cells. It means that
$D(Q,\beta)$-function are defined for all orders.

\section{Summary and discussion}
\setcounter{equation}{0}
We investigated the finite temperature BE correlations of
identical particles in the multiparticle production using the
solutions of the operator field Langevin-type equation in
$S(\Re_{4})$, the quantum version of the Nyquist theorem and the
quantum statistical methods. The model considered states
that all the particles are produced directly from the
a high-energy collision process.
 There was presented the crucial
role of the model in describing the BE correlations via calculations
of DF's as functions of the mean multiplicity and
chaoticity at each four-momentum $\sqrt{Q^2_{\mu}}$.
Based on this model, one can compare the effects
on single particle spectra and multiparticle distribution caused
by multiparticle correlations. There are
several parameters in the model: $\beta, \delta y, \alpha
(\alpha^{\prime})$. One can focus on the statement that the
deviation of the $D(Q,\beta)$-function from zero at finite
values of the physical variables $q^2$ and the model parameter
$\alpha$ indicates that the MPPR should be considered as the
phase-space consisting of the elementary cells (with the size
$\delta y$ of each) which are occupied by the particles of
identical statistics. The Boltzmann behaviour of $R$-function is
available only at large enough values of $ \alpha$ which means
the leading role of the random source contribution to the distribution
function.
An important feature of the model is getting the
information on the space-time structure of the multiparticle
production region. We are able to predict the source size and the
intercept parameter -the chaoticity $\lambda$\ as well.
We have found that DF $R(Q,\beta)$ depends on the number of
elementary defined by the size $\delta y$ in the rapidity
$y$-space.

Of course, the best check of any model could be done if
various kinds of high energy experimental data on the
multiparticle correlations would be well reproduced by the model
in consideration. Our model can be applied to the data which are
available from the ALEPH experiment [10] at $\sqrt{s}$=91.2 GeV.
The ALEPH data applied to $R(Q)$-function (in the notation of
our model) in the region $0.1 GeV \leq Q\leq 1 GeV$ have an essential
nonmonotonic behaviour. It means that the sign of the slope
parameter of $R(Q)$ changes in this region, and the values of
$R(Q)$ are less than unity. In accordance with our model, this
effect could be interpreted as the fact that in the domain
$0.1 GeV \leq Q\leq 1 GeV$ the mean multiplicity $\langle
N(s)\rangle$ has a small value which gives the strong
suppression for the $R(Q)$-function normalized to $\xi(\langle
N(s)\rangle)$. In the case of higher energies of colliding
particles (Tevatron, LHC), the expected (from our model) averaged
 multiplicity
should be very large $\sim$ 300-400, and the nonmonotonic effect
will not occur. We hope that experiments at the Tevatron and LHC
will measure the observables like $R(Q)$ more precisely and the
puzzle regarding the nonmonotonic behavior of the multiparticle
distribution will be clarified.


\end{document}